\documentclass[twocolumn,showpacs,preprintnumbers,floatfix,
               superscriptaddress]{revtex4}

\usepackage{pstcol}
\usepackage{graphicx, fancybox}

  \newcommand{\sls}[1]{\not{\hspace{-2pt}#1}}

\begin{document}

\title{Stable gapless superconductivity at strong coupling}

\author{Masakiyo Kitazawa}
\email{masky@yukawa.kyoto-u.ac.jp}
\affiliation{Institut f\"ur Theoretische Physik, 
J.W.\ Goethe-Universit\"at, D-60438 Frankfurt am Main, Germany}
\affiliation{Yukawa Institute for Theoretical Physics,
Kyoto University, Kyoto 606-8502, Japan}

\author{Dirk H.\ Rischke}
\email{drischke@th.physik.uni-frankfurt.de}
\affiliation{Institut f\"ur Theoretische Physik, 
J.W.\ Goethe-Universit\"at, D-60438 Frankfurt am Main, Germany}

\author{Igor A.\ Shovkovy}
\email{shovkovy@th.physik.uni-frankfurt.de}
  \altaffiliation[on leave from ]{%
       Bogolyubov Institute for Theoretical Physics,
       03143, Kiev, Ukraine}
\affiliation{
Frankfurt Institute for Advanced Studies, 
J.W.\ Goethe-Universit\"at, D-60438 Frankfurt am Main, Germany}

\begin{abstract}

We study cross-flavor Cooper pairing in a relativistic system of 
two fermion species with mismatched Fermi surfaces. We find that 
there exist gapless phases which are characterized by either one 
or two gapless nodes in the energy spectra of their quasiparticles. 
An analysis of the current-current correlator reveals that, at 
strong coupling, both of these gapless phases can be free of 
magnetic instabilities and thus are stable. This is in contrast 
to the weak-coupling case where there are always two gapless 
nodes and the phase becomes magnetically unstable. 

\end{abstract}

\date{\today}
\maketitle


\section{Introduction}  \label{intro}

In recent years, the interest in degenerate fermionic systems
has considerably increased. In part, this was driven by a
substantial progress in experimental studies of trapped cold
gases of fermionic atoms \cite{cold-atoms}. By making use of
various techniques, it has become possible to prepare atomic 
systems of different composition, temperature, density, and 
coupling strength of the interaction. Because of such a 
flexibility, the basic knowledge gained in these studies is 
likely to be of immense value also outside the realm of atomic 
physics. For example, the knowledge of the ground state of an 
asymmetric mixture of two atomic species can shed light on the 
physical properties of strongly interacting dense quark matter 
that may exist in stars.

It is conjectured that the baryon density in the central 
regions of compact (neutron) stars is sufficiently high 
for crushing nucleons (and strange baryons if there are 
any) into deconfined quark matter. The ground state of 
such matter is expected to be a color superconductor 
\cite{old,sc}. (For reviews on color superconductivity, 
see Ref.~\cite{reviews}.) At high density the energetically 
preferred phase is the so-called color-flavor-locked (CFL) 
phase \cite{CFL}. At densities of relevance for compact stars,
however, the situation is much less clear. 

The difficulties in predicting
the ground state of dense quark matter are related to 
the fact that the conditions of charge neutrality
and $\beta$ equilibrium in a macroscopic bulk of matter
such as the core of a compact star have a disruptive effect
on quark Cooper pairing \cite{no2sc,no2sc-other}. For example,
as emphasized in Ref.~\cite{SH03}, enforcing charge neutrality 
can result in unconventional forms of superconductivity, such 
as the gapless 2-flavor color-superconducting (g2SC) phase. 
A 3-flavor version, the so-called gapless color-flavor-locked 
phase (gCFL), was also proposed \cite{gCFL}. While it was argued 
that both types of gapless phases are (chromo-)magnetically unstable 
\cite{HS04,instability-CFL,Fuku05} (for the non-relativistic 
case, see Ref.~\cite{WuYip}), there exist convincing 
arguments that unconventional Cooper pairing in one form or 
another is unavoidable \cite{RajSch}. By taking into account 
the general observation of Refs.~\cite{pd1,pd2,pd3,pd4,pd5} 
that the role of gapless phases diminishes with increasing 
coupling strength, one may naively conclude that gapless 
phases do not exist in the regime of strong coupling. In 
this paper, we show that this conclusion is premature: not 
only do such phases exist, but they can even be magnetically 
stable.

Since the QCD coupling becomes strong at low densities,
one may conjecture that quark matter undergoes Bose-Einstein 
condensation (BEC) of diquarks rather than forming the usual 
Cooper pairs of the Bardeen-Cooper-Schrieffer (BCS) type 
\cite{CSC-BEC,KKKN04,NA05,NNY05}. If this 
is indeed the case, one may observe well-pronounced 
diquark-pair fluctuations in the vicinity of the critical 
temperature \cite{KKKN02,Vosk04,JZ05} and/or the formation 
of a pseudogap phase \cite{KKKN04}. Therefore, it is of 
interest to study Cooper pairing in the strongly coupled 
regime in more detail. 

In this paper, we address the issue of the existence of 
stable gapless phases in a strongly coupled system of two
species of massive fermions. (For a related study, based 
on an effective low-energy description, see also
Ref.~\cite{SS05}). To keep the discussion as simple and 
as general as possible, we consider a model with a local 
interaction that describes two fermion species with equal 
masses, but with non-equal chemical potentials. We find 
that gapless phases with either one or two 
effective Fermi surfaces can exist at strong coupling. 
These phases are stable in the sense that they are free 
of (chromo-)magnetic instabilities \cite{HS04}. It is 
expected that many results of our analysis should remain 
qualitatively similar also in the more complicated case 
of non-equal masses.

The paper is organized as follows. In the next section, we
briefly introduce the model and set up the notation. In
Sec.~\ref{phase diagram}, we present the zero-temperature 
phase diagram in the plane of the average chemical 
potential and the coupling constant for several values of the 
mismatch between the fermion chemical potentials. This diagram, 
while obtained in the mean-field (MF) approximation and, 
thus, not completely reliable at strong coupling, reveals 
a very interesting feature: it has regions of
gapless phases that are free of the Sarma
instability \cite{Sarma} {\em without\/} imposing the neutrality 
condition. Note that this is drastically different from 
the situation at weak coupling, where the absence of 
such an instability is mainly due to charge neutrality 
\cite{SH03,gCFL}, or other types of constraints on the system 
\cite{bp}. In Sec.~\ref{meissner}, we calculate the screening 
masses, defined by the long-wavelength limit of the static 
particle-number current-current correlator,
and reveal a region of parameters (generally at strong
coupling) for which gapless superconductivity is stable.
The discussion of the results and conclusions are
given in Sec.~\ref{conclusion}.

\section{Model and quasiparticle spectrum}
\label{model}

Let us start by introducing the Lagrangian density of the 
model,
\begin{eqnarray}
{\cal L}
=\bar{\psi} (i\sls{\partial} -m +\hat{\mu}\gamma^0 ) \psi
+ {\cal L}_I,
\end{eqnarray}
where $\psi$ denotes the Dirac field which has two internal 
degrees of freedom, called ``flavors'' in the following. In 
general, the mass is a diagonal matrix of the form $\hat m 
= {\rm diag}( m_1, m_2 )$. For simplicity, however, we restrict 
ourselves to the case of equal fermion masses, i.e., we use 
$m_1=m_2=m$ in this paper. No constraints on the values of 
the chemical potentials of the two flavors of fermions are 
imposed, i.e., $\hat\mu = {\rm diag} (\mu_1,\mu_2)$ where 
$\mu_1$ and $\mu_2$ need not be equal.

In order to study superfluidity/superconductivity that 
results from Cooper pairing of different flavors of 
fermions, we introduce the following local interaction 
term to the Lagrangian density,
\begin{eqnarray}
{\cal L}_I
= G (\bar{\psi} i\gamma_5 \sigma_1 C \bar\psi^T )
(\psi^T i\gamma_5 \sigma_1 C \psi),
\label{eq:L_I}
\end{eqnarray}
where $G$ is the coupling constant, $\sigma_1$ is the (symmetric) 
Pauli matrix in flavor space, and $C=i\gamma_0\gamma_2$ is the 
charge conjugation matrix. This term describes a cross-flavor 
attractive interaction between fermions that can drive the 
formation of spin-zero (and, therefore, totally antisymmetric) 
Cooper pairs at weak coupling, or even 
cause the appearance of localized bound states at strong 
coupling. At sufficiently low temperature, these bosonic 
states should form a condensate in the ground state. The 
explicit structure of the condensate is given by the 
following expression,
\begin{eqnarray}
\Delta = 2G \langle \psi^T C i\gamma_5 \sigma_1 \psi \rangle.
\label{Delta}
\end{eqnarray}
In the MF approximation, the value of $\Delta$ 
is determined by the minimization of the effective potential
\begin{eqnarray}
V(\Delta)= \frac{ \Delta^2 }{4G}
- \frac T2 \sum_n \int \frac{ d^3k }{ (2\pi)^3 }
{\rm Tr}\log [ S^{-1}(K) ],
\end{eqnarray}
where $S^{-1}(K)$ is the inverse fermion propagator in 
Nambu-Gor'kov space, 
\begin{eqnarray}
[S(K)]^{-1} = 
\left( \begin{array}{cc}
\sls{K}+\hat\mu \gamma^0 - \hat{m} & -i \sigma_1 \gamma_5 \Delta \\ 
-i \sigma_1 \gamma_5 \Delta & \sls{K}-\hat\mu \gamma^0 - \hat{m}
 \end{array} \right).
\label{eq:propagator}
\end{eqnarray}
The corresponding Nambu-Gor'kov spinor is defined by
$\Psi^{T} = ( \psi^{T}, \psi_C^{T})$ with $\psi_C\equiv 
C {\bar \psi}^T$ being the charge-conjugate spinor.

The (eight) poles of the determinant $\mbox{det}\left[S(K)\right]$ 
determine the dispersion relations of (eight) quasiparticles. In 
the case of equal fermion masses, these can be given explicitly in 
analytical form,
\begin{eqnarray}
k_0=\pm\left(\epsilon_{\pm} \pm \delta\mu \right) , 
\label{eq:spectrum}
\end{eqnarray}
where all eight sign combinations are possible. In the above 
equation, we used the following notation: 
\begin{eqnarray}
\epsilon_\pm &=& \sqrt{ ( E_k \pm \bar\mu )^2 + \Delta^2},\\
E_k &=& \sqrt{ k^2 + m^2 },\\
\bar\mu &=& \frac{\mu_1+\mu_2}{2},\\ 
\delta\mu &=& \frac{\mu_1-\mu_2}{2}
\end{eqnarray}
(without loss of generality, 
we assume that $\bar\mu \geq 0$ and $\delta\mu \geq 0$). 

\begin{figure}[tb]
\begin{center}
\includegraphics[width=0.48\textwidth]{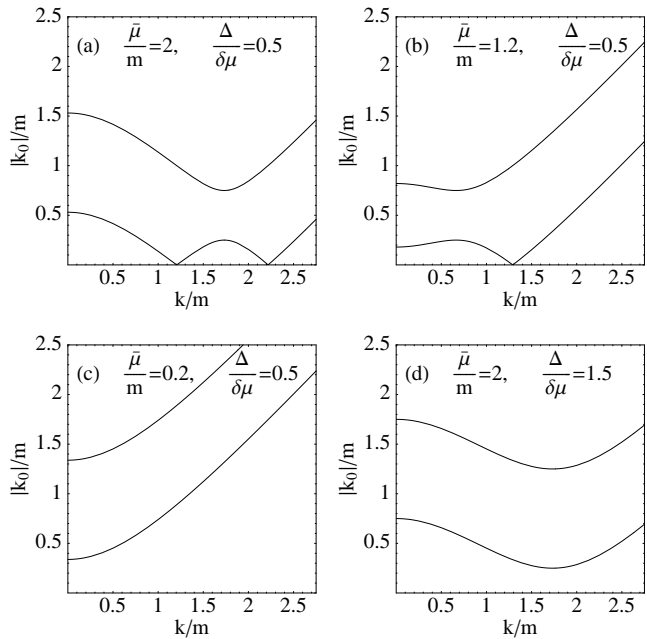} 
\caption{
Typical quasiparticle excitation spectra in gapless and regular 
superconducting phases. The values of $\bar\mu/m$ and $\Delta/\delta\mu$ 
are specified inside the panels. In all four cases $\delta\mu/m=0.5$.
Excitation spectra of antiparticles are not shown.}
\label{fig:spectra}
\end{center} 
\end{figure}

Let us first clarify the structure of the quasiparticle excitation
spectrum keeping $\Delta/\delta\mu$ and $\bar\mu/m$ as free 
parameters, i.e., without actually solving a gap equation.
It is easy to see that, if $\Delta/\delta\mu>1$, there are no 
gapless excitations irrespective of the value of $\bar\mu/m$. 
This is not always the case when $\Delta/\delta\mu\leq 1$, cf. 
Ref.~\cite{SH03}. In this case, gapless modes may exist around 
effective Fermi surfaces at momenta
\begin{equation}
 k_{\pm} = 
\sqrt{\left(\bar\mu\pm\sqrt{\delta\mu^2-\Delta^2}\right)^2-m^2}.
\label{eff_Fermi}
\end{equation} 
Of course, $k_{\pm}$ has to be real, otherwise the corresponding 
effective Fermi surface does not exist. If $\bar\mu/m$ is sufficiently 
large, both $k_{+}$ and $k_{-}$
are real, and there are two effective Fermi surfaces. For $\Delta\neq 0$,
we call the corresponding gapless superconducting phase the gSC(2) 
phase. Such gapless phases were discussed in the context of dense
quark matter \cite{SH03,gCFL,pd1,pd2,pd3,pd4,pd5}. A typical excitation 
spectrum is depicted in Fig.~\ref{fig:spectra}(a). For smaller values of
$\bar\mu/m$, $k_{-}$ has a non-zero imaginary part while $k_{+}$ is real, 
and there is only one effective Fermi surface, cf.\ Fig.~\ref{fig:spectra}(b). We call 
the corresponding gapless superconducting phase the gSC(1) phase.
Finally, for even smaller values of $\bar\mu/m$, both $k_{-}$ and $k_{+}$ 
have non-zero imaginary parts, and there are no gapless excitations in the spectrum; 
we are in a regular, gapped superconducting (SC) phase, cf.\ Fig.~\ref{fig:spectra}(c), 
just as in the case $\Delta/\delta\mu>1$, cf.\ Fig.~\ref{fig:spectra}(d). 

\begin{figure}[tb]
\begin{center}
\includegraphics[width=0.48\textwidth]{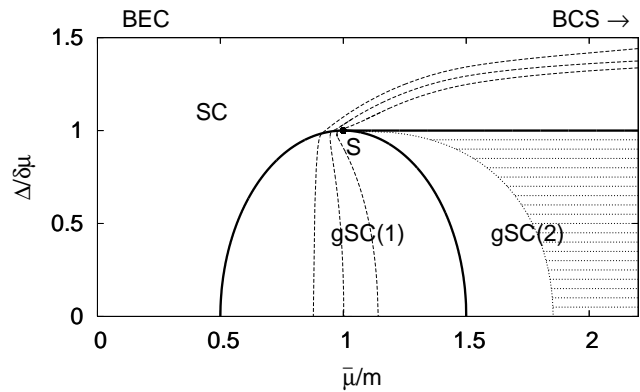} 
\caption{
The phase diagram in the plane of $\Delta/\delta\mu$ and 
$\bar\mu/m$ plotted for a fixed value of $\delta\mu/m=0.5$. 
The region SC represents a superconducting phase without 
gapless modes, while gSC(1) and gSC(2) represent phases 
with one and two effective Fermi surfaces, respectively. 
The three regions merge at the splitting point ``S'', cf. 
Ref.~\cite{SS05}. For a given $m/\Lambda$, the region 
to the right of the dashed line (from top to bottom, 
$m/\Lambda=0.3$, $0.2$, and $0.1$) is not accessible 
in the MF analysis of Fig.~\ref{fig:mu-G_T0}, see text
for a detailed explanation. For $m/\Lambda=0.2$, the 
``screening mass'' $m_M$ is imaginary in the shaded area.}
\label{fig:del-mu}
\end{center} 
\end{figure}

In Fig.~\ref{fig:del-mu}, we show where the three different types 
of superconducting phases occur in the plane of $\Delta/\delta\mu$
and $\bar\mu/m$, for a fixed value of $\delta\mu/m=0.5$. The boundary 
of the region gSC(1) can be derived from the requirement that $k_{-}=0$, 
i.e., the region exists for values of $\Delta/\delta\mu$ and $\bar{\mu}/m$ 
satisfying the condition
\begin{equation}
\left(\frac{\Delta}{\delta\mu} \right )^2 
+\left( \frac{\bar{\mu}-m}{\delta\mu}\right )^2 \leq 1.
\end{equation}
The boundary between the gSC(2) and SC regions is given by 
$\Delta/\delta\mu=1$ for $\bar\mu/m\geq 1$. The three regions 
SC, gSC(1) and gSC(2) meet at the ``splitting'' point S, cf. 
Ref.~\cite{SS05}. 

\section{Phase diagram} 
\label{phase diagram}

In this section, we explore the phase diagram of the model
at hand in the plane of the average chemical potential $\bar\mu$
and the coupling constant $G$. We shall restrict our study to 
the zero-temperature case when the problems with the stability
of gapless phases are expected to be most prominent. Then, the
effective potential reads
\begin{eqnarray}
V(\Delta)&=&\frac{\Delta^2}{4G}
-\int_{0}^{\Lambda} \frac{k^2 d k}{2\pi^2}\Big(
 |\epsilon_{-} + \delta\mu| + |\epsilon_{-} - \delta\mu|
\nonumber \\
&&+|\epsilon_{+} + \delta\mu| + |\epsilon_{+} - \delta\mu| \Big),
\end{eqnarray}
where $\Lambda$ is a momentum cut-off.

\begin{figure}[tb]
\begin{center}
\includegraphics[width=0.48\textwidth]{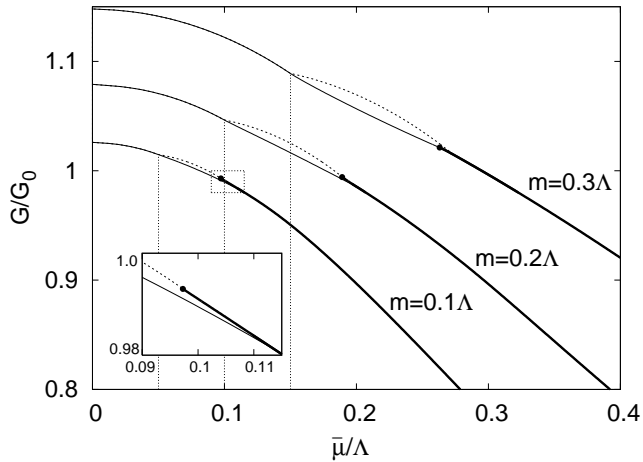} 
\caption{
The phase diagram in the plane of $G/G_0$ and $\bar\mu/\Lambda$
for three different values of the fermion masses, $m/\Lambda=0.1$, 
$0.2$ and $0.3$, and a fixed value of $\delta\mu / m = 0.5$.
The bold and thin lines represent first- and second-order
transitions, respectively. The insertion is an enlarged view 
of the endpoint of the first-order transition line in the case 
of $m/\Lambda=0.1$.}
\label{fig:mu-G_T0}
\end{center} 
\end{figure}

In Fig.~\ref{fig:mu-G_T0}, we show the phase diagram for 
three different values of the fermion masses, $m/\Lambda=0.1$, 
$0.2$ and $0.3$, and a fixed value of $\delta\mu / m = 0.5$.
The coupling constant is normalized by $G_0 = 4\pi / \Lambda^2$.
Note that, for $m=0$ and for $G\geq G_0$, the vacuum is 
unstable with respect to BEC of diquarks. The bold and thin 
solid lines represent first- and second-order transitions, 
respectively. For each choice of mass, the part of the 
diagram above the transition lines corresponds to the fully 
gapped superconducting phase. 
The small regions bounded by the solid and dotted lines in the 
middle of the phase diagram  correspond to gapless phases.  
Below the solid lines, the system is in the normal phase.
The normal phase to the left of the thin dashed vertical 
lines at $\bar\mu=m-\delta\mu$ corresponds to the vacuum.

As one can see from Fig.~\ref{fig:mu-G_T0}, at weak coupling 
the phase transition 
between the normal and superconducting phase is of first order.
This is quite natural when there is a fixed mismatch between 
the chemical potentials of pairing 
fermions. With increasing the coupling constant $G$, the critical 
chemical potential $\bar\mu$ becomes smaller and the lines of 
first-order phase transitions terminate at endpoints. It is 
worth mentioning that the endpoints lie completely inside the 
region of the superconducting phase when $m/\Lambda=0.1$ and 
$0.2$. In the case $m/\Lambda=0.1$, one can see this more 
clearly from the insertion in the lower left corner of 
Fig.~\ref{fig:mu-G_T0}. In the case $m/\Lambda=0.3$, however, 
the corresponding endpoint lies on the phase boundary between 
normal and superconducting matter, or so 
close to it that our numerical resolution is not sufficient 
to make a distinction.

Across the boundary of a second-order phase transition, the gap 
$\Delta$ must be continuous. This means that the gap assumes
arbitrarily small values just above the thin solid lines in 
Fig.~\ref{fig:mu-G_T0}. When the mismatch between the chemical potentials 
$\delta\mu$ is non-zero, there inevitably exists a region adjacent to 
the transition line where $\Delta<\delta\mu$. From Fig.~\ref{fig:del-mu}
we see that this is a necessary condition to have a gapless phase,
and this is precisely what we observe in Fig.~\ref{fig:mu-G_T0}.
It is, however, not a sufficient condition: if $\bar\mu/\Lambda$
is sufficiently small, we could also have a gapped phase, as
seen to the left of the gSC(1) region in Fig.~\ref{fig:del-mu}.

The gapless phases in Fig.~\ref{fig:mu-G_T0} correspond to 
global minima of the effective potential and, thus, they are 
free of the Sarma instability \cite{Sarma}. This might be 
surprising since we do not impose additional constraints
such as neutrality.  An 
apparent discrepancy between this finding and that of 
Ref.~\cite{SH03} is resolved by noting that the stable 
gapless phases in Fig.~\ref{fig:mu-G_T0} always occur at 
strong coupling.

We now want to clarify which regions in Fig.~\ref{fig:del-mu}
are accessible by the MF calculation that gives rise to the 
phase diagram in Fig.~\ref{fig:mu-G_T0}. We first note that 
the region below the solid lines in Fig.~\ref{fig:mu-G_T0} 
represents the vacuum or the normal-conducting phase. Since 
there $\Delta = 0$, moving along its upper boundary corresponds 
to moving along the horizontal axis in Fig.~\ref{fig:del-mu}. 
Across a first-order phase transition, the gap is discontinuous. 
Therefore, certain values of $\Delta/\delta\mu$ are excluded. 
These can be found by computing the values of the gap along both
sides of the bold solid lines in Fig.~\ref{fig:mu-G_T0}. For 
fixed $m/\Lambda$, such a path starts at the merging point of 
the second- and first-order transition lines in Fig.~\ref{fig:mu-G_T0}, 
continues along the lower side of the bold line, and runs 
around the endpoint, before continuing along the upper side of
the first-order transition line. In Fig.~\ref{fig:del-mu},
this path is shown as a dashed line. The region to the right
of this line is not accessible in the MF analysis. As a 
consequence, the only gapless phases appearing in the phase 
diagram in Fig.~\ref{fig:mu-G_T0} are those of type gSC(1). 
This excludes, therefore, possible ground 
states that correspond to the gSC(2) region as well as 
the splitting point S. In fact, the statement regarding 
the splitting point can be made rigorous by noting that the 
second derivative of the effective potential $\partial^2 V/
\partial\Delta^2$ is negative at S, meaning that this point 
cannot be a minimum of $V(\Delta)$. Of course, this 
conclusion may easily change if an additional constraint 
(e.g., such as neutrality) is imposed on the system.

Now, it is natural to ask whether the gSC(1) and gSC(2) types
of gapless phases are subject to the chromomagnetic instability 
\cite{HS04}. This is studied in detail in the next section.

\section{Stability Analysis} 
\label{meissner}

In this section, we discuss the stability of the gapless phases
that were introduced in Sec.~\ref{model}, see Fig.~\ref{fig:del-mu}. 
To this end, we calculate the fermion-number current-current 
correlator and study when such a correlator points toward an 
instability. Note that the use of the fermion-number current 
is not accidental here. When the vacuum expectation value in 
Eq.~(\ref{Delta}) is non-zero, the fermion-number symmetry is 
spontaneously broken.

By definition, the current-current correlator is given by
\begin{equation} 
\Pi^{\mu \nu} (P) = \frac{T}{2}\sum_n\int 
\frac{d^3 {\mathbf k}}{(2\pi)^3} 
{\rm Tr}\left[ \hat{\Gamma}^\mu
{\cal S} (K) \hat{\Gamma}^\nu {\cal S}(K-P) \right] ,
\label{eq:Pi^munu}
\end{equation}
where $\hat\Gamma^\mu\equiv\mbox{diag}(i\gamma^\mu,-i\gamma^\mu)$ 
is the vertex in Nambu-Gor'kov space. Following the approach 
of Ref.~\cite{HS04}, we consider $\Pi^{\mu \nu}$ only in the static 
($p_0=0$) and long wave-length limit ($p\to 0$). Then, it is 
convenient to introduce the ``screening masses'' which are 
defined by
\begin{eqnarray}
m_D^2 &=& -\lim_{{\bf p}\to 0 } \Pi^{00}( \omega=0,{\bf p} ), \\
m_M^2 &=& - \frac12 \lim_{{\bf p}\to 0 }
( g_{ij} + \hat{\bf p}_i \hat{\bf p}_j ) \Pi^{ij}( \omega=0,{\bf p} ).
\label{eq:m_M^2}
\end{eqnarray}
If the fermion-number symmetry is promoted to a gauge symmetry, 
the quantities $m_D$ and $m_M$ would describe the electric and 
magnetic screening properties of a superconductor.

Our calculations show that $m_D^2$ is positive definite in 
the whole of plane of $\Delta/\delta\mu$ and $\bar\mu/m$, 
see Fig.~\ref{fig:del-mu}. However, the magnetic screening 
mass $m_M$ could be imaginary (i.e., $m_M^2<0$) in some cases
when $\delta\mu$ is nonzero. In general, an imaginary result 
for $m_M$ indicates an instability with respect to the formation 
of inhomogeneities in the system \cite{RR,GR05}. 
In the weak-coupling limit, the instabilities develop 
when $\Delta<\delta\mu$ \cite{HS04}. In this paper, we 
study whether a similar instability also develops at 
strong coupling.

When $\Delta$ is nonzero, the Nambu-Gor'kov propagator 
of fermions has both diagonal and off-diagonal components, 
see Eq.~(\ref{eq:propagator}). Their contributions to the 
current-current correlator (\ref{eq:Pi^munu}) can be considered 
separately. Then, the expression for $m_M^2$ can be given in 
the following form:
\begin{eqnarray}
m_M^2 = (m_M^2)_{\rm diag} + (m_M^2)_{\rm off}.
\label{eq:m_M_do}
\end{eqnarray}
Each of the two types of contributions in this equation can 
be further subdivided into particle-particle (pp), 
antiparticle-antiparticle (aa), and particle-antiparticle 
(pa) parts, i.e.,
\begin{eqnarray}
(m_M^2)_{\rm diag} 
= (m_M^2)_{\rm diag}^{\rm (pp)} + (m_M^2)_{\rm diag}^{\rm (aa)} 
+ (m_M^2)_{\rm diag}^{\rm (pa)},
\end{eqnarray}
and a similar representation holds for $(m_M^2)_{\rm off}$. 
The explicit expressions for all contributions are given 
by
\begin{widetext}
\begin{eqnarray}
(m_M^2)_{\rm diag}^{\rm (pp),(aa)}
&=&
\frac13 \int \frac{ d^3k }{(2\pi)^3} \frac{k^2}{E_k^2}
\left(
-\frac{ \Delta^2 }{ \epsilon_\mp^3 } 
\left[ 1 - \theta( -\epsilon_\mp +\delta\mu ) \right]
- \frac{ \epsilon_\mp^2 + E_\mp^2 }{ \epsilon_\mp^2 }
\delta( \epsilon_\mp -\delta\mu ) \right),
\label{eq:m_M_diag^pp}
\\
(m_M^2)_{\rm off}^{\rm (pp),(aa)}
&=&
\frac13 \int \frac{ d^3k }{(2\pi)^3} \frac{k^2}{E_k^2}
\left(
\frac{ \Delta^2 }{ \epsilon_\mp^3 }
\left[ 1 - \theta( -\epsilon_\mp +\delta\mu ) \right]
- \frac{ \Delta^2 }{ \epsilon_\mp^2 }
\delta( \epsilon_\mp -\delta\mu ) \right),
\label{eq:m_M_off^pp}
\\
(m_M^2)_{\rm diag}^{\rm (pa)}
&=&
\frac43 \int \frac{ d^3k }{(2\pi)^3}
\left( 3 - \frac{k^2}{E_k^2} \right)
\left\{
\frac1{ \epsilon_{-}^2 - \epsilon_{+}^2 }
\left(
\frac{ \epsilon_{+}^2 - E_+ E_- }{ \epsilon_{+} }
\left[ 1 - \theta( -\epsilon_{+} +\delta\mu ) \right]
- ( \epsilon_{+} \to \epsilon_{-} )  \right) 
- \frac1{E_k} \right\},
\label{eq:m_M_diag^pa}
\\
(m_M^2)_{\rm off}^{\rm (pa)}
&=&
\frac43 \int \frac{ d^3k }{(2\pi)^3}
\left( 3 - \frac{k^2}{E_k^2} \right)
\frac{\Delta^2}{ \epsilon_{-}^2 - \epsilon_{+}^2 }
\left(
\frac1{ \epsilon_{+} }
\left[ 1 - \theta( -\epsilon_{+} +\delta\mu ) \right]
- ( \epsilon_{+} \to \epsilon_{-} )  \right),
\label{eq:m_M_off^pa}
\end{eqnarray}
\end{widetext}
where $E_\pm = E_k \pm \delta\mu$. The upper and lower signs in 
Eqs.~(\ref{eq:m_M_diag^pp}) and (\ref{eq:m_M_off^pp}) denote the
(pp) and (aa) parts, respectively. It should also be noted that the 
vacuum contribution $1/E_k$ was subtracted in Eq.~(\ref{eq:m_M_diag^pa}).

Several remarks are in order regarding the expressions in 
Eqs.~(\ref{eq:m_M_diag^pp}) through (\ref{eq:m_M_off^pa}). 
Firstly, we note that the contributions due to the terms 
with the $\theta$- and $\delta$-functions in the integrands 
vanish in the gapped superconducting phase (SC). These 
are nontrivial, however, in the gapless phases when there 
exists at least one well-defined effective Fermi surface, 
see Eq.~(\ref{eff_Fermi}). Secondly, the terms with the
$\delta$-function give negative contributions that 
diverge as $-\left[(\delta\mu)^2 - \Delta^2\right]^{-1/2}$ 
when $\Delta \to \delta\mu$ from below. This is a 
reflection of the divergent density of states at $k=k_\pm$ 
(see, e.g.,  the second paper in Ref.~\cite{SH03}). Finally,
we point out a qualitative difference between the integrands 
in the (pp) and (aa) expressions and those in the (pa) ones 
when $k\to 0$. Because of the overall factor $k^2/E_k^2$ in 
the former and the factor $(3-k^2/E_k^2)$ in the latter, the 
low-momentum contributions of the (pp) and (aa) parts are 
suppressed, while those of the (pa) parts are not. This has 
important consequences at strong coupling.

\begin{figure}[tb]
\begin{center}
\includegraphics[width=0.48\textwidth]{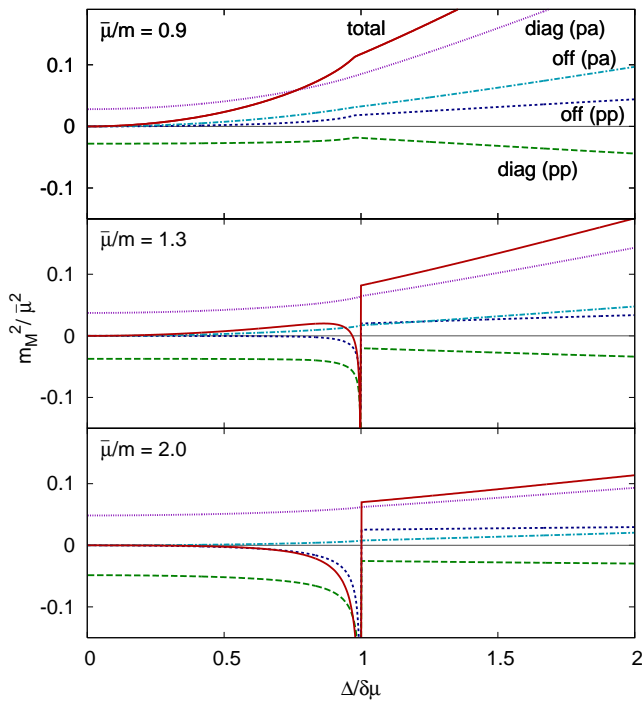} 
\caption{
The dependence of $m_M^2$ (solid line) as well as the (pp) and 
(pa) contributions, defined in Eqs.~(\ref{eq:m_M_diag^pp}) 
through (\ref{eq:m_M_off^pa}), versus $\Delta/\delta\mu$ for 
three different values of $\bar\mu/m$ and a fixed value of 
$m/\Lambda=0.2$.}
\label{fig:mass}
\end{center} 
\end{figure}

Our numerical results for $m_M^2$ as a function of 
$\Delta/\delta\mu$ for three different values of the ratio 
$\bar\mu/m$ (i.e., $0.9$, $1.3$ and $2$) and a fixed value 
of the mass, $m/\Lambda=0.2$, are shown in Fig.~\ref{fig:mass}.
(Note that the qualitative features of the numerical results 
are robust when the parameter $m/\Lambda$ changes in a 
relatively wide range.) The solid lines represent the 
complete results for the screening masses squared, $m_M^2$, 
while the other four lines give the separate (pp) and (pa) 
contributions, defined in Eqs.~(\ref{eq:m_M_diag^pp}) 
through (\ref{eq:m_M_off^pa}). We do not show the (aa) 
contributions in Fig.~\ref{fig:mass} because they are 
always small numerically. 

In each panel of Fig.~\ref{fig:mass}, the value of $m_M^2$ 
vanishes in the limit $\Delta/\delta\mu\to0$ which corresponds 
to the normal phase. As one can infer from the figure, this 
is due to the exact cancellation of the negative (paramagnetic) 
$(m_M^2)_{\rm diag}^{(pp)}$ and positive 
(diamagnetic) $(m_M^2)_{\rm diag}^{(pa)}$ contributions.

As one can see from the top panel of Fig.~\ref{fig:mass}, the result 
for $m_M^2$ is positive definite in the case $\bar\mu/m=0.9$. This
is not so, however, when the value of the ratio $\bar\mu/m$ is larger
than 1. In particular, the quantity $m_M^2$ develops a negative 
divergence just below $\Delta/\delta\mu=1$ and stays negative for 
a range of values of $\Delta/\delta\mu$, see the middle and bottom 
panels in Fig.~\ref{fig:mass}. It is easy to figure out that the 
divergence is caused by the singular behavior of the density of states 
around the effective Fermi surfaces in the gapless phases. This 
singularity affects only the particle-particle contributions, i.e.,
$(m_M^2)_{\rm diag}^{(pp)}$ and $(m_M^2)_{\rm off}^{(pp)}$.
When the value of $\bar\mu/m$ becomes larger than 1 and increases 
further, negative values of $m_M^2$ first appear only near 
$\Delta/\delta\mu=1$ (see, e.g., the middle panel in Fig.~\ref{fig:mass}).
When the ratio $\bar\mu/m$ gets larger than about $1.8$, 
however, the quantity $m_M^2$ becomes negative in the whole 
range $\Delta/\delta\mu<1$ (see, e.g., the bottom panel 
in Fig.~\ref{fig:mass}).

The numerical results for the screening mass can be 
conveniently summarized in Fig.~\ref{fig:del-mu} by identifying 
the region in which $m_M$ is imaginary (i.e., $m_M^2<0$). 
For the given set of parameters, $\delta\mu/m=0.5$ and 
$m/\Lambda=0.2$, this is marked by the shaded area there.
For $\bar\mu/m\gtrsim 1.8$, the gSC(2) type gapless phase
is unstable in the whole range $\Delta/\delta\mu<1$ where
it is defined. This is in agreement, of course, with the 
results at weak coupling \cite{HS04}. It is most 
interesting, however, that for smaller values of the ratio 
$\bar\mu/m$, the quantity $m_M^2$ could be positive even 
in the gapless phases. In particular, $m_M^2>0$ in the 
whole gSC(1) region as well as in a part of the gSC(2) 
region. Note that the gapless phases found in the 
diagram in Fig.~\ref{fig:mu-G_T0} always correspond 
to the stable region.

To complete the analysis of the stability of gapless phases
at strong coupling, we note that a general requirement is 
that the eigenvalues of the susceptibility matrix, 
$-\partial^{2}V/\partial\mu_i\partial\mu_j$, are non-negative
\cite{PWY05}. As a criterion for stability, the susceptibility 
matrix is meaningful only in the ground state defined by that 
$\Delta(\bar\mu,\delta\mu)$ which solves 
the gap equation. In a sense, therefore, the susceptibility 
criterion has a less general status than the one of the 
(chromo-)magnetic stability. The latter does not refer to a 
specific form of the gap equation. This difference might be 
very important when additional constraints (e.g., neutrality) 
are enforced on the system.

In the MF approximation used in this study, the eigenvalues 
of the susceptibility matrix are non-negative in all phases 
of Fig.~\ref{fig:mu-G_T0}, as 
well as in the regions to the left of the dashed lines in 
Fig.~\ref{fig:del-mu}. We argue that this is related directly 
to the fact that the ground state is defined as the global 
minimum of the effective potential. Indeed, one of the 
important necessary conditions for the susceptibility 
criterion to be satisfied reads \cite{PWY05}
\begin{equation}
-\frac{\partial^2 V}{\partial(\delta\mu)^2}\geq0.
\end{equation}
In the model at hand, as one can easily check, this 
is equivalent to 
\begin{equation}
\frac{\partial^2 V}{\partial\Delta^2}\geq 0,
\end{equation}
which is always satisfied at the global minimum of the effective
potential. 

Before concluding this section, we would like to briefly discuss
the properties at the splitting point S in Fig.~\ref{fig:del-mu}.
This point describes the situation when the two effective Fermi 
surfaces merge exactly at zero momentum. This coincides
with the definition in Ref.~\cite{SS05} where the properties 
near S are studied using an effective theory approach. From
Fig.~\ref{fig:del-mu} we see that four qualitatively different 
regions merge at point S, i.e., (i) gSC(1), (ii) SC, (iii) gSC(2) 
with a positive value of $m_M^2$, and (iv) gSC(2) with a negative 
value of $m_M^2$. This topology is in agreement with the 
qualitative picture presented in Ref.~\cite{SS05}, although
here we use a different approach. It is interesting, though, 
that the ground state defined by the splitting point 
is never realized in the MF approximation in our model.
This is because the second derivative of the thermodynamic
potential is negative definite at the point S, which excludes 
the possibility of having a minimum there.

\section{Discussion and Summary} 
\label{conclusion}

In this paper, we explored the superconducting (superfluid) 
phases in a relativistic model with two fermion species 
having mismatched Fermi surfaces. We found that, in general,
there could exist gapless phases with either one or two 
effective Fermi surfaces. In the MF approximation, however,
only the gapless phase gSC(1) is realized as a ground 
state, and only at strong coupling. We also calculated the 
fermion-number current-current correlator in the static and 
long wave-length limit. The results show that the gapless 
phases at strong coupling are free of the (chromo-)magnetic 
instability. This is in contrast to the situation in neutral 
quark matter at weak coupling \cite{HS04}. 

It is interesting to note that there could exist different 
regimes of BEC in the model at hand. In general, bound 
bosonic states are formed when $\bar\mu/m<1$ \cite{NSR85}. 
As seen from Fig.~\ref{fig:del-mu}, this is satisfied in 
parts of the SC and gSC(1) regions, giving rise to gapped 
and gapless phases, respectively. While the number densities 
of the two species of fermions are equal in the former, there 
is an excess of one of the species in the latter. In fact, 
this excess is an order parameter that defines the gapless 
phases at zero temperature \cite{SH03}. The gapless BEC phase
is a mixture of tightly-bound bosons in the form of a 
condensate and additional unpaired fermions \cite{SS05,PWY05}. 

It is worth mentioning that the gapless phase of the gSC(1) type 
could also exist for a range of parameters when $\bar\mu/m>1$.
In this case, stable bosons do not exist and, thus, the BEC 
regime cannot be realized. If the diquark coupling in quark matter 
is sufficiently strong, this gapless phase can potentially 
be realized as the ground state of baryon matter. Our analysis 
of the one-loop fermion contribution to the current-current 
correlator suggests that the gSC(1) phase is stable. In the 
case of quark matter, however, it might be important to check 
if the inclusion of gluon and ghost contributions does not 
change the conclusion. While such contributions are negligible 
in the weak-coupling limit, this may not be the case at strong 
coupling.

A few words are in order regarding non-relativistic models 
\cite{SS05,PWY05,SheehyRadzihovsky,KunYang,SachdevYang}. In our 
analysis, we saw that purely relativistic effects due to the 
antiparticle-antiparticle loops were always small near the 
splitting point. This suggests strongly that our results should 
remain qualitatively the same also in the non-relativistic limit 
that is defined by $|\bar\mu-m|/m\ll 1$, $\delta\mu/m\ll 1$ and 
$\Delta/m\ll 1$. Indeed, these conditions define a narrow area 
around the vertical line $\bar\mu/m=1$ in Fig.~\ref{fig:del-mu} 
which includes the point S. In fact, this also suggests that the 
topology around the splitting point is the same in both relativistic 
and non-relativistic models.

{\em Note added.} While finishing this paper, we learned that 
a similar study in a non-relativistic model is done by E.~Gubankova, 
A.~Schmitt, and F.~Wilczek \cite{GSW06}.

\section*{Acknowledgments}

M.K. thanks T.~Kunihiro for encouragements. 
I.A.S. acknowledges discussions with Andreas Schmitt.
The work of M.K. was supported by Japan Society for 
the Promotion of Science for Young Scientists. 
The work of D.H.R. and I.A.S. was supported in part by 
the Virtual Institute of the Helmholtz Association 
under grant No. VH-VI-041, by the Gesellschaft f\"{u}r 
Schwerionenforschung (GSI), and by the Deutsche 
Forschungsgemeinschaft (DFG).

\end{document}